\documentstyle[psfig, rotate]{aa}

\begin{document}

\title{Clues to the origin of parsec to kilo-parsec jet 
misalignments in EGRET sources}
\author{Xinwu Cao\inst{}}
\institute{  {\inst{1} Shanghai Astronomical Observatory, Chinese Academy of Sciences, 80 Nandan Road,
Shanghai, 200030, China}\\
{\inst{2} National Astronomical Observatories, Chinese Academy of Sciences,
China} \\
{\inst{3} Beijing Astrophysical Center (BAC), Beijing, China} }

\date{Received ; accepted December 21, 1999}
\thesaurus{03(11.01.2; 11.10.1; 11.17.2; 02.01.2)}

\offprints{cxw@center.shao.ac.cn}

\maketitle
\markboth
{X. Cao: Clues to the origin of parsec to kilo-parsec jet 
misalignments}
{X. Cao: Clues to the origin of parsec to kilo-parsec jet 
misalignments}

\begin{abstract}

The apparent position angle difference $\Delta$PA between parsec and 
kilo-parsec jets in blazars can be related to the bending properties of the jets. 
We present  correlations  between the misalignment $\Delta$PA and the
ratio of radio to broad-line emission for a sample of $\gamma$-ray
blazars. The present study is limited to EGRET sources due to 
uniform data for the radio and optical properties of these sources being
easily available. 
The broad-line emission is known to be a good indicator of the accretion power
for both steep and flat-spectrum quasars.
We find significant correlations between $\Delta$PA and the ratio
of radio flux (total and radio core flux respectively) to
broad-line flux, which is consistent with the beaming
scenario. A weak correlation between $\Delta$PA and the ratio
of extended radio to broad-line flux might imply that the
intrinsic bend of the jet is related with the ratio of jet mechanical
power to accretion power.

\end{abstract}

\begin{keywords}
galaxies:active-- galaxies:jets-- quasars:emission lines-- accretion,
accretion disks
\end{keywords}

\section{Introduction}

The recent released third catalog of high-energy $\gamma$-ray sources by 
the EGRET telescope on the Compton Gamma-Ray Observatory contains 66 
high-confidence identifications of blazars and 27 lower-confidence 
potential blazar identifications (Hartman et al. 1999). The $\gamma$-ray 
blazars are relativistically beamed and have strong jet components. The 
inverse Compton scattering is the most promising process responsible 
for the $\gamma$-ray emission (Marscher \& Gear 1985; Bloom \& Marscher 1996; 
Ghisellini \& Madau 1996; B\"ottcher \& Dermer 1998). All $\gamma$-ray 
AGNs identified 
by EGRET are radio-loud with flat spectrum, but not all flat-spectrum 
AGNs are detectable $\gamma$-ray sources. One possibility is that the 
beaming cone for $\gamma$-ray emission is narrower than that for radio 
emission. In this case, the $\gamma$-ray emission could be beamed away 
from the line of sight, but the radio emission is still Doppler boosted 
due to a wider beaming cone (von Montigny et al. 1995).  

The apparent position angle difference $\Delta$PA between parsec and 
kilo-parsec jets can be related to the bending properties of the jets.  
Many authors have investigated the misalignment angle distributions 
for the different samples of radio sources (Pearson \& Readhead 1988; 
Conway \& Murphy 1993; Appl et al. 1996; Tingay et al. 1998). 
If the angle between the relativistic jet and the line of the sight
is small, the projection effect of beaming can amplify the intrinsic
bend of the jet (Conway \& Murphy 1993). 
Several effects such as motion of the host galaxy, collision of the
jet with clouds, or precession of the central engine, can thereby
cause the bend.

Recently, the concept of jet-disc symbiosis was introduced and the
inhomogeneous jet model plus mass and energy conservation in the jet-disc 
system was applied to study the relation between disc and
jet luminosities (Falcke \& Biermann 1995; Falcke et al. 1995;
Falcke \& Biermann 1999).
An effective  approach to study the link between
these two phenomena is to explore the relationship between luminosity
in line emission and kinetic power of jets in different
scales (Rawlings \& Saunders 1991; Celotti, Padovani \& Ghisellini 1997).
Rawlings \& Saunders (1991) derived the total jet kinetic power $Q_{jet}$
and found a correlation between $Q_{jet}$ and the narrow
line luminosity $L_{NLR}$. Similar  correlations between line and radio
emission have also been found by some authors (Cao \& Jiang 1999;
Xu et al. 1999; Willott et al. 1999).
The optical line region is photoionized by
a nuclear source (probably radiation from the disk), so
the optical line emission is a better accretion power indicator than 
the optical continuum radiation may be enhanced by
relativistically beamed synchrotron radiation for some
flat-spectrum quasars (Celotti et al. 1997). The extended
radio emission which is not
affected by beaming may reflect the jet power $Q_{jet}$.  
Thus, the ratio of extended radio to broad-line flux
reflects the ratio of jet power to accretion power. 
In this work, we present  correlations  between the misalignment
$\Delta$PA and the ratio of radio to broad-line emission for a sample
of $\gamma$-ray blazars.
In Sect. 2, we describe the sample of sources.
The results are contained in Sect. 3. The last section includes 
the discussion.

\begin{table*}[htb]
 \begin{flushleft}
  \caption{Data of radio and broad-line fluxes.}
  \begin{tabular}{ccccllrrrl}\hline
Source & Class. & z & log f$_{BLR}$ & Lines & Refs. & f$_{5{\rm G}}$(Jy)
& $\alpha_{11-6}$ & f$_{ext}$(mJy) & Refs.\\
(1) & (2) & (3) & (4) & (5) &(6) & (7) & (8) & (9)& (10)\\ \hline
 0119+041$^\dagger$& Q &0.637 & -12.61& Mg\,{\sc ii}, H$\gamma$, H$\beta$&JB91,RS80 & 1.67 & 0.04 & 32.9 & BM87\\
 0234+285$^\dagger$& Q &1.210 & -12.66& C\,{\sc iv} & W84& 1.44 & -0.24& 9.75  & BM87 \\ 
 0235+164& BL& 0.940 &  -13.79& Mg\,{\sc ii} & S93a  & 2.85 & 1.03& 11.5 & M93 \\ 
 0336$-$019& Q &0.852 & -12.55 & Mg\,{\sc ii}, H$\gamma$, H$\beta$ &B89,JB91 & 2.86 & 0.30& 49.7 & BM87 \\ 
 0420$-$014& Q &0.915 & -12.70& Mg\,{\sc ii} &B89 & 1.46 & 0.01& 8.23 & BM87 \\ 
 0440$-$003& Q &0.844 & -13.00& Mg\,{\sc ii}, H$\gamma$, H$\beta$ &B89,JB91 & 2.61 & -0.29& 32.15 & BM87 \\ 
 0458$-$020& Q &2.286 & -13.28& Ly$\alpha$, C\,{\sc iv} & B89& 1.74 & -0.08& 121.8 & BM87 \\ 
 0521$-$365$^\dagger$& BL& 0.055 &  -11.80& Ly$\alpha$, H$\beta$, H$\alpha$ & S95 & 9.29 & -0.43& 3900 & AU85 \\ 
 0537$-$441 &BL & 0.896 &  -12.55& Mg\,{\sc ii}  & S93a  & 4.00 & 0.06& 10.95 & BM87 \\ 
 0539$-$057$^\dagger$& Q & 0.839 & -13.42 &  Mg\,{\sc ii} & SK93 & 1.55 & 1.41& ... &  \\
 0804+499$^\dagger$& Q &1.433 & -12.71& C\,{\sc iv}, Mg\,{\sc ii} &L96 & 2.05 & 0.47& 1.84 & M93 \\
 0836+710& Q &2.172 & -12.12& Ly$\alpha$, C\,{\sc iv}, Mg\,{\sc ii} & L96& 2.59 & -0.32& 33.5 & M93 \\ 
 0851+202& BL &0.306 & -12.88& Mg\,{\sc ii}, H$\beta$, H$\alpha$& S89,S93a& 2.62 & 0.11&1.87  & K92 \\ 
 0954+658& BL &0.367 & -14.04& H$\alpha$ & L96& 1.46 & 0.35& 8.32 & K92 \\ 
 1101+384 & BL & 0.031 &-12.94 &  H$\alpha$ & M92 & 0.73 & -0.09& 72.8 & AU85 \\ 
 1127$-$145$^\dagger$ & Q& 1.187 &  -12.13& Ly$\alpha$ & W95  & 6.57 & 0.03&40.9  & BM87 \\ 
 1156+295 & Q & 0.729 & -12.49 & Mg\,{\sc ii}& W83 & 1.54 & -0.1& 111.4 & AU85 \\
 1222+216& Q & 0.435 &  -12.11& H$\beta$ & SM87 & 1.26 & -0.40& 156.9 & H83 \\
 1226+023& Q &0.158 & -10.27& Ly$\alpha$, C\,{\sc iv}, H$\beta$, H$\alpha$ &O94,JB91,Ma96 &42.85 & 0.15& 4003 & BM87 \\ 
 1253$-$055& Q &0.536 & -12.42& Ly$\alpha$, C\,{\sc iv}, Mg\,{\sc ii}, H$\beta$, H$\alpha$ & W95,Ma96,N79&14.95 & 0.30& 502.5 & BM87 \\ 
 1334$-$127 & Q & 0.539 & -12.88 & Mg\,{\sc ii} & S93b & 2.24 & 0.17& 109.9 & BP86 \\
 1424$-$418& Q &1.522 & -13.20& Mg\,{\sc ii} & S89& 3.13 & 0.28& ... &  \\
 1510$-$089& Q &0.361 & -12.00& Ly$\alpha$, Mg\,{\sc ii}, H$\gamma$, H$\beta$, H$\alpha$ & N79,T93,O94,BK84 & 3.08 & 0.31& 80.6 & BM87 \\ 
 1514$-$241$^\dagger$ & BL & 0.042 & -13.86 & H$\beta$ & T93 & 2.00 & -0.16& 12.8 & AU85 \\
 1611+343& Q &1.401 & -12.17& Ly$\alpha$, C\,{\sc iv} & W95& 2.67 & 0.10& 40.9 & BM87 \\
 1622$-$253 &Q& 0.786 & -13.80 & Mg\,{\sc ii}, H$\beta$ &d94 &2.08 &-0.14 & ... &  \\
 1633+382& Q &1.814 & -12.52& Ly$\alpha$, C\,{\sc iv}, Mg\,{\sc ii} &L96 & 4.02 & 0.73& 9.7 & BM87 \\ 
 1652+398& BL & 0.0337 & -13.50& H$\alpha$ & L96 & 1.42 & 0.06& 26.9 & AU85 \\
 1725$-$044 & Q & 0.296 & -12.35 & H$\beta$ & R84 & 1.24 & 0.76 & 5.1 &  BM87\\
 1730$-$130 & Q & 0.902 & -12.78& H$\gamma$, H$\beta$ & J84 & 6.99 & 0.8& 42.8 & BM87 \\
 1739+522& Q &1.379 & -12.90& C\,{\sc iv}, Mg\,{\sc ii} & L96& 1.98 & 0.68& 6.8 & BM87 \\ 
 2230+114& Q &1.037 & -11.87& Ly$\alpha$, C\,{\sc iv} & W95& 3.61 & -0.50& 74.4 & BM87 \\ 
 2251+158& Q &0.859 & -11.88& Ly$\alpha$, C\,{\sc iv}, Mg\,{\sc ii}, H$\gamma$, H$\beta$ & N79,W95,JB91&17.42 & 0.64& 386.7 & BM87 \\ 
 2351+456& Q &1.992 & -13.58& C\,{\sc iv}, Mg\,{\sc ii} & L96 & 1.42 & -0.05& ... &  \\
\hline

\end{tabular}
\end{flushleft}
\begin{minipage}{170mm}

Notes for the table 1. Q: quasars; BL: BL Lac objects. 
$\dagger$: lower-confidence blazar identification.\\
Column (1): IAU source name. Column (2): classification of the source.
Column (3): redshift. Column (4): estimated total
broad-line flux (erg s$^{-1}$ cm$^{-2}$). Column (5): lines from which 
the total $f_{line}$ has been estimated. Column (6): references for the line 
fluxes. Column (7): radio flux density (in Jy) at 5 GHz. Column (8): two-point 
spectral index between 6$-$11 cm. Column (9): extended radio flux
density (in mJy) at 5GHz in the rest frame of the source. Column (10): references
for the extended radio flux density. \\

References:

AU85: Antonucci \& Ulvestad (1985),
B89: Baldwin et al. (1989),
BK84: Bergeron \& Kunth (1984), 
BM87: Browne \& Murphy (1987), 
BP86: Browne \& Perley (1986),
d94: di Serego Alighieri et al. (1994),
H83: Hintzen et al. (1983),  
J84: Junkkarinen et al. (1984), 
JB91: Jackson \& Browne (1991), 
K92: Kollgaard et al. (1992), 
L96: Lawrence et al. (1996),
M92: Morganti et al. (1992),
M93: Murphy et al. (1993), 
Ma96: Marziani et al. (1996), 
N79: Neugebauer et al. (1979), 
O94: Osmer et al. (1994),
R84: Rudy (1984), 
RS80: Richstone \& Schmidt (1980), 
S89: Stickel et al. (1989),
S93a: Stickel et al. (1993a),
S93b: Stickel et al. (1993b),  
S95: Scarpa et al. (1995),
SK93: Stickel \& K\"uhr (1993), 
SM87: Stockton \& MacKenty (1987), 
T93: Tadhunter et al. (1993), 
W83: Wills et al. (1983), 
W84: Wampler et al. (1984),
W95: Wills et al. (1995). 

 \end{minipage}

\end{table*}

\begin{table*}[htb]
  \begin{flushleft}
  \caption{Data of radio core flux density and $\Delta$PA.}
  \begin{tabular}{ccccclcclcl}\hline
Source & $\nu_1$ & f$_1$ & $\nu_2$ & f$_2$ & Refs. & PA$_{kpc}$ & PA$_{pc}$ 
& Refs. & $\Delta$PA & ref. \\
(1) & (2) & (3) & (4) & (5) &(6) & (7) & (8) & (9) & (10) & (11)\\ \hline
0119+041$^\dagger$ & 2.29 & 1.0 & 8.4 & 0.53 & P85,M86 & 146 & 100 & P82,K98 &46 &  \\
0234+285$^\dagger$ & 2.32 & 1.78 & 8.55 & 1.32 & F96 & -25 & -10 & M93,F96& 15 & H98 \\
0235+164 & 5 &  4.45 & 22 & 3.44 & S97,Mo96 & -37 & 45 & M93,M90 & 82 & H98 \\
0336$-$019 & 2.29 & 1.4 & 22 & 2.08 & P85,Mo96 & -15 & 9 & P82,W92 & 24 & A96 \\
0420$-$014 & 2.32 & 3.1 & 8.55 & 1.19 & F96 & 180 & -147 & AU85,F96 & 33 & H98  \\
0440$-$003 & 2.32 & 2.11 & 8.55 & 0.85 & FC97 & 90 & 102 & P82,FC97 & 12 & H98 \\
0458$-$020 & 5 & 2.62 & 22 & 0.82 & S97,Mo96 & -126 & -55 & B87,W92 & 71 & A96  \\
0521$-$365$^\dagger$ & 4.8 & 1.2 & 8.4 & 1.65 & T98 & 305 & 310 & K86,T96 & 5& H98   \\
0537$-$441 & 4.851 & 4.0 & 8.4 & 2.55 & T96,M86 & 305 & 5 & P82,T96 & 60 & H98 \\
0539$-$057$^\dagger$& 2.32 & 0.77 & 8.55 & 0.89 & FC97 & 50 & 149 & P82, FC97 & 99 & \\
0804+499$^\dagger$ &  5 & 1.344 & 22 & 0.54 & PR88,Mo96 & 199 & 116 & CM93& 83  & CM93   \\
0836+710 & 1.7 & 1.11 & 5 & 1.045 & H92,PR88 & 200 & 214 & P82,PR88 & 14 & H98\\
0851+202 & 2.32 & 1.23 & 8.55 & 1.6 & FC97 & 253 & 246 & R87 & 7 & K92  \\
0954+658 & 2.29 & 0.43 & 5 & 0.48 & P85,G92 & 206 & -65 & P82,M90 & 90 & CM93\\
1101+384 & 1.7 & 0.224 & 5 & 0.366 & P95,X95 & 301 & 316 & ZB90,T98 & 15 & T98 \\
1127$-$145$^\dagger$ & 5 & 1.91 & 8.387 & 1.4 & S97,T98 &41 & 65 & R88,W92& 24 & H98  \\
1156+295 & 2.32 & 0.94 & 8.55 & 1.00 & F96 & 60 & 24 & M90,T98 & 36 & T98 \\
1222+216 & 2.29 & 0.33 & 5 & 0.36 & P85,Ho92b & ... & ... & ... & 23 & Ho92a  \\
1226+023 & 5  & 21.8 & 22 & 7.79 & S98,Mo96 & 222 & -130 & P82,W90 & 8 & H98   \\
1253$-$055 & 2.3 &  3.97 &  8.4 & 2.12 & M87,L90 & 202 & -137 & P82,U87 & 21& M90  \\
1334$-$127 & 2.32 & 3.45 & 8.55 & 4.27 & F96 & 90 & 155 & P82,F96 & 65 & \\
1424$-$418 & 2.29 & 0.37 & 5 & 1.36 & P85,S98 & 56 & -10 & T98 & 66 & T98  \\
1510$-$089 & 2.32 & 2.97 & 8.55 & 1.56 & FC97 & 160 & 173 & P82,R84 & 13 & H98  \\
1514$-$241$^\dagger$ & 5 & 1.53 & 22 & 0.41 & S98,Mo96 & 120 & 177 & P82,S98 & 57 & \\
1611+343 & 2.32 &  4.16 & 8.55 & 2.14 & FC97 & -170 & -178 & M93,F96 & 8 & H98  \\
1622$-$253 & 2.32 & 1.67 &  8.55 & 1.7 & FC97 & 303 & 6 & P82,F96 & 63 & H98  \\
1633+382 & 2.32 & 2.41 & 8.55 & 1.03 & FC97 & -60 & -64 & X95,PR88 & 4 & H98  \\
1652+398& 2.32 & 0.69 & 8.55 & 0.49 & FC97 & 45 & 110 & P95,X95 & 65 & T98 \\
1725$-$044 & 2.32 & 0.84 & 8.55 & 0.51 & FC97 & 105 & 113 & P82,FC97 & 8 & \\
1730$-$130 & 5 & 2.34 & 8.387 & 8.8 & S97,T98 & 0 & 273 & T98 & 87 & T98  \\
1739+522 & 2.32 & 0.94 & 8.55 & 0.44 & FC97 &260 & 319 & P82,PR88 & 59 & A96  \\
2230+114 & 2.32 & 2.11 & 8.55 & 1.2 & FC97 & 140 & 137 & P82,B87 & 3 & M90  \\
2251+158 & 2.29 & 3.4 & 5 & 0.9 & P85,M87 &-49 & -65 & B82,P84 & 16 & B87   \\
2351+456 & 5 & 0.323 & 22 & 0.42 & PR88,Mo96 & 25 & 321 & T98 & 64 & T98  \\
\hline
\end{tabular}

\end{flushleft}
\begin{minipage}{150mm}

Notes for the table 2. $\dagger$: lower-confidence blazar identification.\\
Column (1): IAU source name. Column (2)-(4): VLBI observation frequences
(in GHz) and core flux density(in Jansky).
Column (6): references for core flux density. Column (7) and (8):
position angles (degree) for kilo-parsec and parsec jet components.
Column (9): references for position angles.
Column (10) and (11): difference of position angle (degree) between
kilo-parsec and parsec jet components and the reference in which the
$\Delta$PA is given.\\

References:

A96: Appl et al. (1996), 
AU85: Antonucci \& Ulvestad (1985),  
B82: Browne et al. (1982), 
B87: Browne (1987), 
CM93: Conway \& Murphy (1993), 
F96: Fey et al. (1996), 
FC97: Fey \& Charlot (1997), 
G92: Gabuzda et al (1992), 
H92: Hummel et al. (1992), 
Ho92a: Hooimeyer et al. (1992a),
Ho92b: Hooimeyer et al. (1992b),
H98: Hong et al. (1998), 
K86: Keel (1986), 
K92: Kollgaard et al. (1992), 
K98: Kellermann et al. (1998), 
L90: Linfield et al. (1990), 
M86: Morabito et al. (1986), 
M87: Madau et al. (1987), 
M90: Mutel (1990), 
M93: Murphy et al. (1993), 
Mo96: Moellenbrock et al. (1996), 
P82: Perley (1982), 
P84: Pauliny-Toth et al. (1984), 
P85: Preston et al. (1985), 
P95: Polatidis et al. (1995), 
PR88: Pearson \& Readhead (1988), 
R84:  Romney et al. (1984),  
R87: Roberts et al. (1987), 
R88: Rusk (1988), 
S97: Shen et al. (1997), 
S98: Shen et al. (1998), 
T96: Tingay et al. (1996), 
T98: Tingay et al. (1998), 
U87: Unwin (1987),  
W90: Wardle et al. (1990), 
W92: Wehrle et al. (1992), 
X95: Xu et al. (1995). 
ZB90: Zhang \& Baath (1990). 

\end{minipage}
\end{table*}

\section{The sample}

Complete information on the line spectra is available for very few 
sources in our sample, since different lines are observed for the sources
at different redshifts.
We have to estimate the total broad-line flux from the available observational
data. There is not a solidly established procedure to derive the total
broad-line flux and we therefore adopt the method proposed by Celotti et al. 
(1997). The following lines: Ly$\alpha$, C\,{\sc iv},  Mg\,{\sc ii}, 
H$\gamma$, H$\beta$ and H$\alpha$, which contribute the major parts in 
the total broad-line emissions, are used in our estimate. We use the line 
ratios reported by Francis et al. (1991) and add the contribution from 
line H$\alpha$ to derive the total
broad-line flux (see Celotti et al. 1997 for details).
We then search the literature to collect data on broad-line fluxes.
We only consider values of line fluxes (or luminosities) given directly
or the equivalent width and the continuum flux density  at the corresponding line
frequency which are reported together in the literature. When more than one
value of the same line flux was found in the literature, we take the most
recent reference.

We start with the $\gamma$-ray blazars identified by Hartman et al.(1999) 
including lower-confidence blazars.
There are 79 AGNs with available redshifts in the third EGRET catalog. 
Among these sources, we search
the literature extensively and find 44 sources with sufficient line data
to estimate the total broad-line flux. The remainder of the sources that
lack broad-line flux data include 9 BL Lac objects and 26 quasars. The
broad-line fluxes have not been measured due to weak line emission for the
BL Lac objects. The situation for the quasars is quite different from the
BL Lac objects. We note that the spectroscopic observations for most of
these 26 quasars have been performed. However, the line data of these
quasars are usually incomplete, i.e., only the equivalent width, line
profile or line-to-continuum ratio is given, but the continuum flux
density at the given frequency is not available probably due to the specific purpose
of the literature or the problem of calibration. Only a bit more than half
of the $\gamma$-ray sources with known redshifts have sufficient data,
such that the total broad-line flux can be estimated. This is
similar to the situation in Cao \& Jiang (1999). In their work, 198
sources within the starting sample of 378 sources have suitable data to
derive the total broad-line flux.
No evidence shows that the lack of broad-line flux
for these sources would affect the main results of present analyses, though
it leads to a highly incomplete sample for present study.  
The further spectroscopic observations on these sources would be helpful. 
We collect the data of all sources with both the broad-line flux and the
misalignment angle $\Delta$PA between
parsec and kilo-parsec jets, which leads to a sample of 34 blazars
(we add the TeV $\gamma$-ray objects: Mkn501 to the sample, which is not
listed in the EGRET catalog).
There are 26 quasars and 8 BL Lac objects in this sample,  
in which 7 sources are lower-confidence potential blazar identifications
and two TeV $\gamma$-ray objects: Mkn421 and Mkn501. The broad-line data
are listed in Table 1. We compile the data of the extended radio
flux density 
at 5 GHz in the rest frame of the source in column (9) of Table 1. The data
given at the wavelength other than 5 GHz are K-corrected to 5 GHz in the
rest frame of the source assuming $\alpha_{\rm ext}=-0.75$
($f_{\rm ext}\propto\nu^{\alpha_{\rm ext}}$).

We also give  the core flux density data at two different
frequencies for each source
in the sample, and a two-point spectral index is then derived for the core
of the source. The core flux density at 5 GHz in the rest frame
of the source is
available by K-correction. The misalignment between kilo-parsec and parsec
jets $\Delta$PA are taken from the literature. For a few sources,
different values of $\Delta$PA are given by different authors usually due
to the complex jet structures. We take the minimum $\Delta$PA for these
sources. All data of the core flux density  and the misalignment angle $\Delta$PA
are given in table 2.

The misalignment angle $\Delta$PA is
given by the comparison between the VLA and VLBI maps. We note that
only one position angle (VLA or VLBI) is available for some
sources. Only the VLA position angle is available for the sources
0414$-$189 and 0954+556. There are five sources: 0454$-$234, 1504$-$166, 1741$-$038,
2200+420 and 2320$-$035, of which only the VLBI position angle is available.
One reason is that these sources are too compact.
High dynamic range VLA maps of many sources sufficient to reveal
weak kilo-parsec structure are not available. 
Therefore, further radio observations on these sources
might reveal their misalignment information. Further
spectroscopic observations are also necessary to complete the sample.

\begin{figure}
\centerline{\psfig{figure=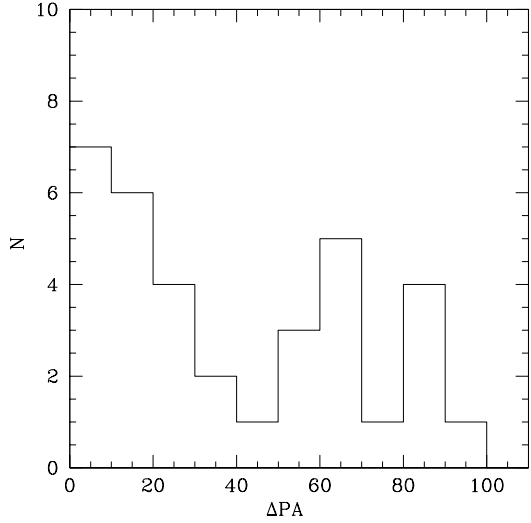,width=8.0cm,height=8.0cm}}
\caption{The $\Delta$PA distribution of the sample. }
\end{figure}

\begin{figure}
\centerline{\psfig{figure=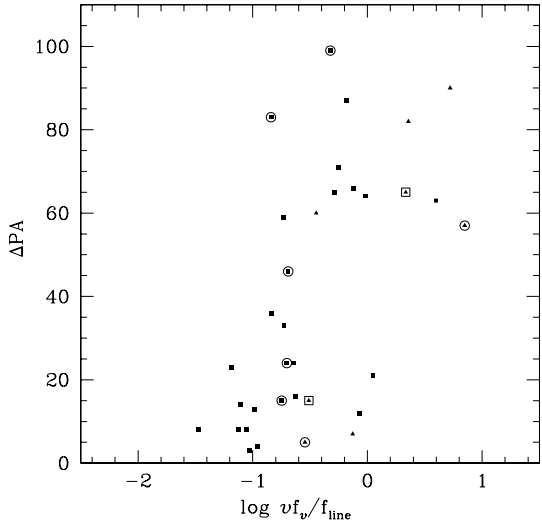,width=8.0cm,height=8.0cm}}
\caption{The relation between the misalignment $\Delta$PA and the ratio
of radio to the broad-line flux.
The full squares represent the quasars, and the full triangles represent
the BL Lac objects, while large circles denote
the lower-confidence blazar identifications by EGRET and large squares
denote the TeV $\gamma$-ray objects. }
\end{figure}

\begin{figure}
\centerline{\psfig{figure=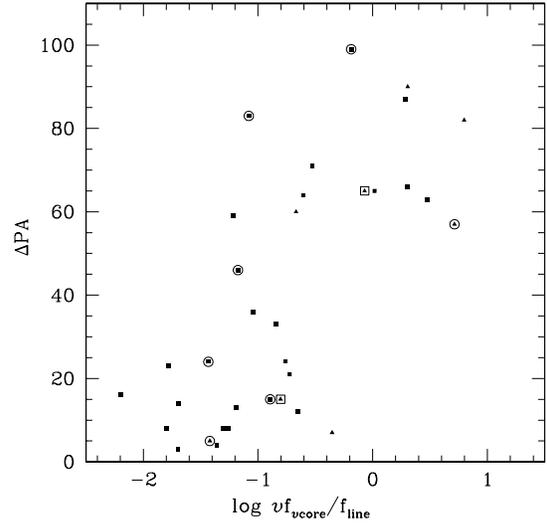,width=8.0cm,height=8.0cm}}
\caption{The relation between the misalignment $\Delta$PA and the ratio
of VLBI core flux to the broad-line flux (symbols as in Fig. 2.). }
\end{figure}

\begin{figure}
\centerline{\psfig{figure=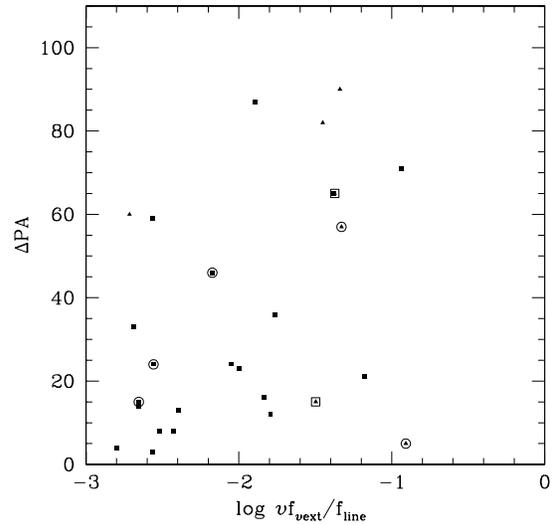,width=8.0cm,height=8.0cm}}
\caption{The relation between the misalignment $\Delta$PA and the ratio
of the extended radio flux to the broad-line flux (symbols as in Fig. 2.). }
\end{figure}

\section{Results}

The distribution of $\Delta$PA of the sample is plotted in Fig. 1 and
appears bimodal, which is
similar to that given by Pearson \& Readhead (1988). Such distribution
can be explained by projection of the helical jet (Conway \& Murphy 1993). 
In Fig. 2 we plot the relation between the apparent misalignment $\Delta$PA 
and the ratio of the total radio flux at 5 GHz to the broad-line flux. The radio 
flux density is K-corrected to the rest frame of the source. A correlation is
found at 99.9 per cent significant level for the whole sample using Spearman's
correlation coefficient $\rho$.
A slightly less significant correlation is present for quasars 
in the sample. We also find a significant correlation between $\Delta$PA
and the ratio of the VLBI core flux to the broad-line flux at the level
99.99 per cent, where the VLBI core flux density is also K-corrected to the rest frame
of the source at 5 GHz (see Fig. 3).
In Fig. 4 we plot the relation between the apparent misalignment $\Delta$PA 
and the ratio of the extended radio flux at 5 GHz in the rest frame
of the source to the broad-line flux.
A weak correlation at 90 per cent significance shows that a trend
for sources with higher ratios have larger misalignment $\Delta$PA.

In present sample, there are two TeV $\gamma$-ray sources: Mkn421
(1101+384) and Mkn501 (1652+398). We know that the three TeV
$\gamma$-ray objects are quite different from  other $\gamma$-ray sources
(Coppi \& Aharonian 1999a,b). Only Mkn421 is listed in the third EGRET
catalog (Hartman et al. 1999). However, we cannot find obvious different
behaviours for these two TeV $\gamma$-ray objects from the remains in the
sample (see Figs. 2 $-$ 4, labeled by large squares).

The redshifts of the sources in our sample ranges from 0.031 to 2.286, which
means that the spatial resolution of VLA and VLBI are very different.
However, the angular resolution is about same for these sources, which would
probably affect the correlation analyses (Appl et al. 1996). We therefore
group the sources into low (z$<$0.5) and high (z$>$0.5) redshift objects.
For the latter, the angular to linear scale mapping is approximately
constant. We re-analyze the correlations and find a correlation
between $\Delta$PA and the ratio of the radio core flux to the
broad-line flux at 99.94 per cent significant level for the 24
objects with z$>$0.5.

\section{Discussion}

Recently, Serjeant et al. (1998) found a correlation between 
radio and optical continuum emission for a sample of steep-spectrum radio quasars 
that is evidence for a link between  accretion process and jet power
Their sample is limited to the steep-spectrum quasars to reduce 
the Doppler beaming effect in the optical waveband.
A similar correlation is given by Carballo et al (1999) for
the B3-VLA sample. 
Cao \& Jiang (1999) present a correlation between radio and broad-line
emission for a sample of radio quasars including both flat-spectrum and
steep-spectrum quasars. They adopted the broad-line emission instead of
the optical continuum in their investigation to avoid the beaming effect
on optical emission.
The jet power $Q_{jet}$ is proportional to the bulk
Lorentz factor of the jet, and also depends on the size of the jet and the
particle density in the jet (Celotti \& Fabian 1993).
The apparent misalignment may be affected by the angle between the
direction of VLBI jet and the sight line if the angle is small (Conway \&
Murphy 1993). The total radio flux density and VLBI core flux density
are both strongly increased by beaming in core-dominated radio blazars.
The sources with higher ratio may therefore have higher Doppler factors. 
So, the correlations between $\Delta$PA and $\nu f_{\rm rad}/f_{\rm
line}$ or $\nu f_{\rm core}/f_{\rm line}$ can be explained  by the beaming
effects. 
Impey et al. (1991) found that
the sources with higher fractional optical polarization tend to
have relatively larger misalignment $\Delta$PA between VLBI and parsec structures,
which is consistent with both large apparent misalignments
and optical fractional polarization being correlated with large
beaming and small angles to the line of sight. 

The extended radio emission may also be correlated to the jet power
$Q_{jet}$ (see Fig. 4). If this is true, the ratio of the extended radio flux to
the broad-line flux may reflect the ratio of the jet power to the disk
luminosity: $Q_{jet}/L_{acc}$.
The weak correlation between $\Delta$PA and the ratio
of extended radio to broad-line flux might imply that the
intrinsic bend of the jet is related with the ratio of jet
mechanical power to accretion power.

Jets in quasars may be powered by  rotating
black holes in the nuclei (Blandford \& Znajek 1977; Moderski et al. 1998).
The jet power $Q_{jet}$ is then 
related to the rotational energy of the black hole according to the
BZ mechanism.
Hence, the ratio of radio to broad-line flux might be related to the angular
momentum of the black hole to some extent. Also, we know that
the variation of the rotation axis of the black hole caused by the
Lense-Thirring effect can result in a change of the orientation of the
nuclei jet (Bardeen \& Petterson 1975; Scheuer \& Feiler 1996), which may
lead to the different ejected orientations between small
and large scale jets.
The faster rotating black hole may cause the jet ejected
orientation changing more rapidly, which leads to a larger
intrinsic bend of the jet. 

More recently, Ghosh \& Abramowicz (1997), Livio et al. (1999)
have shown that the electromagnetic output from the inner disc is generally
expected to dominate over that from the hole. If this is the case for
the $\gamma$-ray AGNs, the jet power $Q_{jet}$ from the disc is mainly
determined by the poloidal magnetic field $B_{pd}$ at the disc
surface. A tentative explanation on the relation in Fig. 4. is that
both bending and the jet power could be increased by having large magnetic
fields in the accretion disk.

The further study on this problem for a larger sample of core-dominated
radio quasars will be in our future work, and we can then check
whether the correlations are properties only of EGRET sources or of all
blazars in general.

\acknowledgements {
I thank the referee for his helpful comments and suggestions
that improved the presentation of this paper,
X.Y. Hong is thanked for helpful discussion on the
measurement of misalignment angle. The support by NSFC and Pandeng Plan is
gratefully acknowledged. This research has made use of
the NASA/IPAC Extragalactic Database (NED), which is operated by the Jet
Propulsion Laboratory, California Institute of Technology, under contract
with the National Aeronautic and Space Administration. }

{}

\newpage
\end{document}